\documentclass[twocolumn,trackchanges]{aastex61}

\newcommand\aastex{AAS\TeX}

\usepackage{graphicx}

\submitjournal{ApJ}

\shorttitle{\aastex\ Jet Triggered by Reconnection between Super-penumbral Fibrils and a Mini-filament}
\shortauthors{Yang et al.}

\begin{document}

\title{Observations of a Solar Jet Triggered by Reconnection between Super-penumbral Fibrils and a Mini-filament}

\correspondingauthor{Liheng Yang}
\email{yangliheng@ynao.ac.cn}

\author[0000-0002-0786-7307]{Liheng Yang}
\affil{Yunnan Observatories, Chinese Academy of Sciences, Kunming 650216, China}
\affiliation{Yunnan Key Laboratory of the Solar physics and Space Science, Kunming 650216}

\author[0000-0003-2891-6267]{Xiaoli Yan}
\affiliation{Yunnan Observatories, Chinese Academy of Sciences, Kunming 650216, China}
\affiliation{Yunnan Key Laboratory of the Solar physics and Space Science, Kunming 650216}

\author[0000-0003-4078-2265]{Qingmin Zhang}
\affiliation{Purple Mountain Observatory, Chinese Academy of Sciences, Nanjing 210023, China}

\author[0000-0002-6526-5363]{Zhike Xue}
\affiliation{Yunnan Observatories, Chinese Academy of Sciences, Kunming 650216, China}
\affiliation{Yunnan Key Laboratory of the Solar physics and Space Science, Kunming 650216}

\author[0000-0002-9121-9686]{Zhe Xu}
\affiliation{Yunnan Observatories, Chinese Academy of Sciences, Kunming 650216, China}
\affiliation{Yunnan Key Laboratory of the Solar physics and Space Science, Kunming 650216}

\author[0000-0003-4393-9731]{Jincheng Wang}
\affiliation{Yunnan Observatories, Chinese Academy of Sciences, Kunming 650216, China}
\affiliation{Yunnan Key Laboratory of the Solar physics and Space Science, Kunming 650216}

\author{Fangyu Xu}
\affiliation{Yunnan Observatories, Chinese Academy of Sciences, Kunming 650216, China}
\affiliation{Yunnan Key Laboratory of the Solar physics and Space Science, Kunming 650216}

\author{Roslan Umar}
\affiliation{East Coast Environmental Research Institute (ESERI), Universiti Sultan Zainal Abidin, 21300 Kuala Nerus,Terengganu, Malaysia}

\author[0000-0002-0464-6760]{Yian Zhou}
\affiliation{Yunnan Observatories, Chinese Academy of Sciences, Kunming 650216, China}

\author{Defang Kong}
\affiliation{Yunnan Observatories, Chinese Academy of Sciences, Kunming 650216, China}
\affiliation{Yunnan Key Laboratory of the Solar physics and Space Science, Kunming 650216}

\author{Weijie Meng}
\affiliation{Yunnan Observatories, Chinese Academy of Sciences, Kunming 650216, China}
\affiliation{Yunnan Key Laboratory of the Solar physics and Space Science, Kunming 650216}

\author{Xinsheng Zhang}
\affiliation{Yunnan Observatories, Chinese Academy of Sciences, Kunming 650216, China}
\affiliation{Yunnan Key Laboratory of the Solar physics and Space Science, Kunming 650216}

\author[0000-0003-2045-8994]{Qiaoling Li}
\affiliation{Department of Physics, Yunnan University, Kunming, Yunnan 650091, China}

\author{Liping Yang}
\affiliation{School of Physics, Electrical and Energy Engineering, Chuxiong Normal University, Chuxiong 675000, China}

\begin{abstract}

Coronal jets are highly dynamic phenomena in the solar atmosphere, yet their driving mechanisms remain an active topic of investigation. In this paper, we report a coronal jet triggered by the interaction between super-penumbral fibrils and a mini-filament, based on coordinated observations from the New Vacuum Solar Telescope (NVST), the Chinese H$\alpha$ Solar Explorer (CHASE), and the Solar Dynamics Observatory (SDO). The fibrils were anchored between the negative-polarity region of a sunspot and an emerging positive-polarity region associated with a moving magnetic feature (MMF). As the positive polarity migrated outward, the fibrils elongated and interacted with the mini-filament, one of whose footpoints was rooted in pre-existing negative-polarity fields. Intense brightenings at the interaction site, together with changes in the connectivity of the mini-filament footpoint from the pre-existing negative polarity to the sunspot, indicate the occurrence of magnetic reconnection. The event produced a narrow hot jet accompanied by a broader cool component. The cool plasma exhibited a clockwise rotation, providing evidence for the transfer of magnetic twist during reconnection. Persistent magnetic flux cancellation was observed before and during the jet eruption. These observations demonstrate that small-scale magnetic structures, such as MMFs, can significantly influence mini-filament eruptions and highlight the important role of flux cancellation in triggering coronal jet activity.

\end{abstract}

\keywords{Sun: activity --- Sun: chromosphere --- Sun: corona --- Sun: magnetic fields}

\section{Introduction}

Jets are collimated or slightly curved plasma ejections that occur throughout the solar atmosphere. They are observed across a variety of solar environments, including active regions, quiet Suns, and coronal holes, and in a wide range of wavelengths such as H$\alpha$, extreme ultraviolet (EUV), ultraviolet (UV), X-rays, and white-light \citep[e.g.,][]{1992PASJ...44L.173S,1996ApJ...464.1016C,2002ApJ...575..542W,2010ApJ...720..757M,2014A&A...567A..11Z,2016ApJ...832L...7P,2018ApJ...853..189P,2018ApJ...854...92T,2020A&A...642A..44H,2022A&A...657A.118Q}. Recent high-resolution observations have unveiled rich fine-scale structures within jets, including plasma blobs \citep[e.g.,][]{2019ApJ...870..113Z,2022FrASS...8..238C,2023ApJ...942...86Y}, coexisting cool and hot plasma components \citep[e.g.,][]{1999ApJ...513L..75C,2007A&A...469..331J}, and signatures of instabilities such as the Kelvin-Helmholtz instability \citep[e.g.,][]{2018NatSR...8.8136L,2019ApJ...884L..51Y}. In many cases, jets recur from the same source region, suggesting the presence of sustained or intermittent driving mechanisms \citep[e.g.,][]{2011RAA....11.1229Y,2013A&A...555A..19G,2017Ap&SS.362...10J,2018ApJ...869...39M,2019ApJ...887..220Y,2019ApJ...887..154L,2023ApJ...945...96Y,2024ApJ...962L..38D,2025ApJ...987..193Y}. These jets are ubiquitous features that span multiple atmospheric layers (from the chromosphere to the corona) highlighting their crucial role in solar atmospheric dynamics. A growing body of observational evidence indicates that jets contribute significantly to coronal heating and solar wind acceleration, primarily through their ability to transport mass, momentum, and energy into the upper solar atmosphere \citep[e.g.,][]{2007Sci...318.1591S,2014Sci...346A.315T,2023Sci...381..867C}. Numerous studies have provided detailed reviews of the physical parameters, initiation processes, and theoretical frameworks associated with solar jets \citep{2016AN....337.1024I,2016SSRv..201....1R,2022AdSpR..70.1580S}.

Magnetic reconnection is widely recognized as the primary mechanism responsible for the generation of solar jets. The emerging-flux model was initially proposed to explain coronal jets, in which reconnection occurs between newly 
emerging magnetic bipoles and the ambient open magnetic field lines \citep{1992PASJ...44L.173S,1995Natur.375...42Y}. Typical observational features, such as inverted-Y shaped structures, whiplike motions, transient brightenings 
at the jet base, and high-speed plasma outflow, are consistently associated with jet activity, providing strong evidence for this mechanism \citep{1992PASJ...44L.173S,1996ApJ...464.1016C}. On this basis, \citet{2009ApJ...691...61P} 
proposed the embedded-dipole model, which involves null-point reconnection within a fan-spine magnetic topology. In this scenario, a twisted closed magnetic structure, embedded within an oppositely directed background field, 
reconnects with surrounding untwisted open field lines. This process releases magnetic helicity accumulated in the corona and produces a helical or untwisting jet. In subsequent work, \citet{2010ApJ...720..757M} demonstrated that 
magnetic breakout reconnection, triggered by the eruption of mini-filaments, can also give rise to solar jets. A successful eruption leads to the formation of a blowout jet, characterized by broader and more complex dynamics, whereas 
a failed eruption typically results in a standard jet with simpler morphology \citep{2015Natur.523..437S}. The magnetic breakout reconnection scenario for coronal jets has been further explored through numerical modeling by \citet{2017Natur.544..452W}, providing additional support for its role in jet initiation and evolution. This model has received substantial observational support, further validating its applicability in explaining solar jet 
dynamics \citep[e.g.,][]{2011ApJ...735L..18L,2015Natur.523..437S,2017ApJ...835...35H,2019ApJ...887..239Y}.

Moving Magnetic Features (MMFs) are small, dynamic magnetic elements that migrate outward from sunspots toward the boundary of the surrounding moat region \citep{2003A&ARv..11..153S,2019PASJ...71R...1H,2019ApJ...876..129L}. These features are typically categorized into three types: Type I MMFs, which consist of bipolar pairs; Type II MMFs, which are monopolar features with the same polarity as their parent sunspot; and Type III MMFs, which are monopolar features with the opposite polarity \citep{2000eaa..bookE2038S}. A number of observational studies have established a strong association between MMFs and the formation of solar jets. \citet{2007ApJ...656.1197B} demonstrated that collisions between MMFs and the opposite polarity magnetic field of an emerging flux region can trigger surges, as well as chromospheric and coronal brightenings. Furthermore, \citet{2015ApJ...815...71C} reported that recurrent solar jets at the edges of active regions are driven by continuous magnetic cancellation between MMFs and the pre-existing opposite polarity magnetic field. To gain further insight into the underlying physical mechanisms, \citet{2013ApJ...777...16Y} performed a 2.5-dimensional numerical magnetohydrodynamic simulation to model the magnetic reconnection between MMFs and the surrounding magnetic field, driven by the horizontal motion of magnetic structures in the photosphere. The simulation revealed that, as a result of this reconnection, two distinct types of jets arise: a hot jet generated by Joule heating and a cool jet produced by the sling-shot effect.

The evolution of the photospheric magnetic field, including both magnetic flux emergence and cancellation, is closely associated with the formation and dynamics of coronal jets. However, whether emergence or cancellation serves as the dominant driving mechanism remains under debate. Numerous observational studies suggest that flux emergence plays a key role in jet generation. For example, \citet{1992PASJ...44L.173S} found that many X-ray jets are linked to emerging flux regions, leading to the proposal of the emerging-flux reconnection model. \citet{2016ApJ...833..150L} reported recurrent homologous jets from an emerging flux region, and \citet{2023ApJ...947L..17L} recently identified a multi-thermal jet triggered by flux emergence. This scenario is further supported by two- and three-dimensional numerical simulations \citep[e.g.,][]{1995Natur.375...42Y,2005ApJ...635.1299A,2008ApJ...673L.211M,2009ApJ...691...61P,2015ApJ...798L..10L}. In contrast, growing observational evidence points to magnetic flux cancellation as a crucial mechanism in jet production. \citet{2016ApJ...832L...7P} analyzed 13 quiet-Sun coronal jets and found they were driven by mini-filament eruptions triggered by flux cancellation along the underlying neutral line. Their follow-up study identified similar behavior in coronal-hole jets \citep{2018ApJ...853..189P}, and more recently, \citet{2024MNRAS.528.1094Y} reported an active-region jet also initiated by cancellation. Flux cancellation is thought to facilitate the buildup of sheared and twisted magnetic fields around pre-eruption mini-filaments and to trigger their eruptions \citep{2020ApJ...902....8C}. This mechanism has also been extensively investigated through magnetohydrodynamic (MHD) simulations, which offer theoretical support for the observational findings \citep[e.g.,][]{2024ApJ...960...51P,2025A&A...699A..87P}.

In this study, we utilize high-resolution, multi-wavelength observations from the New Vacuum Solar Telescope \citep[NVST;][]{2014RAA....14..705L,2020ScChE..63.1656Y}, the Solar Dynamics Observatory \citep[SDO;][]{2012SoPh..275....3P}, and the Chinese H$\alpha$ Solar Explorer \citep[CHASE;][]{2022SCPMA..6589602L} to investigate the interaction between a set of chromospheric super-penumbral fibrils and a mini-filament. The fibrils are rooted at one end in a sunspot with negative magnetic polarity and at the other end in an emerging positive polarity of an MMF. Our objective is to explore the physical consequences of these interactions and provide new observational evidence to better understand the underlying processes. The remainder of the paper is organized as follows: Section 2 describes the observational data sets and methods. Section 3 presents the main results, and Section 4 summarizes our conclusions and provides further discussion.

\section{Observations and Data Analysis}
 
The NVST features a 1-meter aperture and operates under vacuum conditions. It is specifically engineered to capture high-resolution observations of the Sun’s photosphere and chromosphere. 
The facility integrates three major subsystems: a multi-channel high-resolution imaging system, a multi-wavelength spectrometer, and a high-order adaptive optics (AO) module. The imaging system enables simultaneous 
observations in multiple wavelengths. It targets the photosphere via the TiO band at 7058~{\AA}, and captures chromospheric details through the H$\alpha$ and He {\sc i} lines at 6563~{\AA} and 10830~{\AA}, respectively. 
For H$\alpha$ observations, a tunable Lyot filter is incorporated, allowing spectral scanning across a $\pm$5~{\AA} window with 0.1~{\AA} step size and a spectral bandwidth of 0.25~{\AA}. The spectrometer operates in two spectral 
windows: Fe {\sc i} 5324~{\AA} and Ca {\sc ii} 8542~{\AA}. The AO system features 151 actuators and achieves a Strehl ratio exceeding 0.75 under favorable seeing conditions ($\leq$10 cm) \citep{2016ApJ...833..210R,2016SPIE.9909E..2CZ,2023SCPMA..6669611Z}. During the observation period from 01:45 to 03:01 UT on 2023 June 5, the NVST focused on active region NOAA 13326. A jet event was detected in the H$\alpha$ 6563~{\AA} line-center, $\pm$0.4~{\AA} off-band, and He {\sc i} 10830~{\AA} images. For H$\alpha$, the observations have a pixel size of approximately 0.165$''$ with a cadence of 42 seconds, and the raw data were processed with dark subtraction, flat-field correction, and speckle reconstruction using the speckle masking technique \citep{2016NewA...49....8X,2022RAA....22f5010C}. In comparison, the He {\sc i} 10830~{\AA} images have a finer pixel size of about 0.145$''$ and a higher cadence of 10 seconds. These data underwent dark and flat-field corrections, followed by reconstruction using the Nonrigid Alignment-based Solar Image Reconstruction method \citep{2022RAA....22i5005L}.
 
The CHASE performs high-resolution spectroscopic observations in the H$\alpha$ waveband. It is equipped with two payloads, including the Solar Atomic Frequency Discriminator and the H$\alpha$ Imaging Spectrograph (HIS). The HIS operates in two distinct observational modes: raster scanning mode (RSM) and continuum imaging mode (CIM). In RSM mode, the instrument acquires solar spectra in the H$\alpha$ (6559.7--6565.9~{\AA}) and Fe {\sc i} (6567.8--6570.6~{\AA}) wavebands with a high spectral resolution of 0.024~{\AA}~pixel$^{-1}$ and a temporal cadence of 1 minute. In this study, we utilize the H$\alpha$ spectral data obtained in RSM mode by the HIS instrument. The data were calibrated from Level 0 to Level 1 through a standard processing pipeline, which includes dark-field and flat-field corrections, slit image curvature correction, wavelength and intensity calibration, and coordinate transformation \citep{2022SCPMA..6589603Q}. The Doppler velocity of the jet was derived using the center-of-gravity method. The centroid wavelength of the observed H$\alpha$ profile was calculated as \( \lambda_0 = \frac{\sum_{i} \lambda_i (I_c - I_i)}{\sum_{i} (I_c - I_i)} \), where $I_c$ is the continuum intensity and $I_i$ is the observed intensity at wavelength $\lambda_i$. The Doppler shift was then determined relative to the theoretical center of the H$\alpha$ line at 6562.82~{\AA}. Corrections were subsequently applied to account for the effects of solar rotation and satellite motion.

The Atmospheric Imaging Assembly \citep[AIA;][]{2012SoPh..275...17L} and the Helioseismic and Magnetic Imager \citep[HMI;][]{2012SoPh..275..207S} are two primary instruments onboard the SDO. AIA continuously captures full-disk solar images in seven extreme ultraviolet (EUV) channels, specifically 94~{\AA}, 131~{\AA}, 335~{\AA}, 211~{\AA}, 193~{\AA}, 171~{\AA}, and 304~{\AA}, as well as in three ultraviolet and visible continuum bands 
centered at 4500~{\AA}, 1700~{\AA}, and 1600~{\AA}. In this study, the differential emission measure (DEM) is reconstructed using near-simultaneous data from six coronal-sensitive AIA EUV bands: 171~{\AA}, 193~{\AA}, 211~{\AA}, 335~{\AA}, 
94~{\AA}, and 131~{\AA} \citep{2012ApJ...761...62C}. The resulting DEM is then used to compute the emission measure (EM), which serves as a diagnostic of the thermal structure associated with the brightening features at the base of the jet. 
HMI contributes full-disk observations of the line-of-sight magnetic field with a spatial resolution of 1$''$ and a cadence of 45 seconds, allowing detailed analysis of photospheric magnetic field evolution at the jet base.

\section{Results}

\subsection{The Source Region of the Jet}

On June 5, 2023, a jet was observed emanating from the western edge of active region NOAA 13323, which exhibited a $\beta\gamma$ magnetic configuration. This active region was located at S07E21. Figure 1 illustrates the magnetic environment associated with the jet. In the HMI line-of-sight magnetogram, an emerging positive-polarity region of an MMF (indicated by the blue arrow in Figure 1(a)) is situated close to a pre-existing negative-polarity patch (marked by the red arrow in Figure 1(a)) within the jet’s source region. The vector magnetic field (see the zoomed-in image in Figure 1(a)) shows that the emerging positive-polarity region is magnetically connected to the negative polarity of the sunspot. To emphasize the magnetic configuration, line-of-sight magnetic field contours derived from SDO/HMI observations are overlaid on the H$\alpha$ image (see Figure 1(c)), with red and blue contours representing positive and negative magnetic polarities, respectively, at a contour level of $\pm$50 G. Notably, one end of a series of super-penumbral fibrils is rooted in the sunspot with negative polarity (labeled by the white “–” in Figure 1(c)), while the other end is anchored at the emerging positive-polarity region of the MMF (labeled by the white “+” in Figure 1(c)). A sunspot typically consists of a dark central umbra surrounded by a filamentary penumbra in the photosphere, which hosts a diverse array of fine-scale dynamic phenomena \citep{2003A&ARv..11..153S,2019PASJ...71R...1H}. This filamentary structure extends into the chromosphere, where H$\alpha$ imaging reveals radially oriented fibrils surrounding the sunspot, resembling a magnified version of the photospheric penumbra. This extended chromospheric structure is referred to as super-penumbral fibrils \citep{1968SoPh....5..489L}, which trace magnetic field lines that span a much broader area of the solar surface than those confined to the photospheric penumbra \citep{2019ApJ...880..143J}. In addition to the fibrils, a prominent dark structure is identified, as indicated by the white arrow in Figure 1(c). The southern portion of this structure is extremely thin, making it difficult to identify it as a typical mini-filament. However, \citet{2024ApJ...960..109S} reported that some active-region jet eruptions are associated with only a comparatively thin mini-filament “strand”. Therefore, this feature may represent a mini-filament, with one footpoint anchored in the pre-existing negative polarity (marked by the pink “–” in Figure 1(c)). It is also evident in the NVST He {\sc i} 10830~{\AA} (marked by a white arrow in Figure 1(d)) and SDO/AIA 193~{\AA} (marked by white arrows in Figures 1(b) and 1(e)) images. Because the structure is not clearly aligned with the magnetic polarity inversion line, its chirality cannot be reliably determined.

\subsection{Magnetic Reconnection between Super-penumbral Fibrils and a Mini-filament}

Figure 2 illustrates the interaction between the super-penumbral fibrils and the mini-filament that led to the formation of a jet. Prior to the interaction, one end of the super-penumbral fibrils was anchored in the sunspot, while the other was rooted in the emerging positive-polarity region of an MMF (see Figure 2(b1)). In contrast, one end of the mini-filament was connected to the pre-existing negative polarity. It is worth noting that between 01:45:53 and 01:57:26 UT, the mini-filament exhibited apparent growth, suggesting that it may have still been undergoing formation during this period (see Animation 1 in the NVST H$\alpha$ and He {\sc i} 10830~{\AA} wavelengths). At approximately 01:57:26 UT, faint brightenings appeared at the base of the jet in the SDO/AIA 1600~{\AA} channel (marked by a black arrow in Figure 2(a2)), corresponding to absorption features in the NVST He {\sc i} 10830~{\AA} image (marked by a blue arrow in Figure 2(c2)). This suggests that the material in the brightening region absorbed background emission, resulting in a local decrease in intensity. At the same time, the connectivity of certain fibrils within the mini-filament was altered. Their footpoints, initially anchored in the negative polarity, shifted to the sunspot (see the white dashed line in Figure 2(b2)). These signatures indicate that magnetic reconnection occurred between the super-penumbral fibrils and the mini-filament. As reconnection progressed, more fibrils of the mini-filament underwent connectivity changes (see the white dashed line in Figure 2(b3)). Around 01:58:17 UT, a bright, narrow jet appeared in the 304~{\AA} channel (marked by a white arrow in Figure 2(d3)), providing further evidence of reconnection. About two minutes later, a dark jet developed to the left of the bright one (see Animation 1), representing the cool component of the jet, while the bright structure corresponded to its hot component. The hot component might be consists of plasma heated at the interface between the super-penumbral fibrils and the mini-filament through Joule dissipation, and subsequently accelerated by the stress released during magnetic reconnection \citep{1995Natur.375...42Y, 2013ApJ...777...16Y}. The cool component likely represents the ejection of cool mini-filament material. As shown in Figure 2(b4), line-of-sight magnetic field contours derived from SDO/HMI observations are overlaid on the H$\alpha$ image, revealing that the connectivity of some fibrils within the mini-filament had shifted to the sunspot penumbra. The ejected cool component of the jet is distinctly visible in the H$\alpha$, He {\sc i} 10830~{\AA}, and 304~{\AA} images, as marked by the yellow arrows in Figures 2(b4)-(d4). 
 
The brightenings at the base of the jet persist for an extended period, with the brightness continuously fluctuating, suggesting ongoing magnetic reconnection. To explore the thermal properties of the brightenings, we constructed emission measure (EM) maps at 01:58 UT and 02:02 UT (see Figures 3(a) and (b)). These brightenings are distinctly visible in the EM maps (indicated by black squares), showing a significant enhancement in emission compared to the surrounding background. The average differential emission measure (DEM) distributions for the brightenings are displayed in Figures 3(c)-(d), derived by subtracting the pre-event DEM at 01:56 UT from the DEM at the corresponding locations. As shown in Figures 3(c)-(d), the DEM distribution spans from Log T = 6.0 (1 MK) to Log T = 7.0 (10 MK), indicating a multi-thermal structure. The peak temperature is found to be Log T = 6.4 (2.5 MK), implying that the brightenings are likely the result of magnetic reconnection.

\subsection{Evolution of the jet}

Evolution of the jet is illustrated in the NVST H$\alpha$ line-center images, NVST H$\alpha$ Dopplergrams, and SDO/AIA 304~{\AA} and 131~{\AA} images shown in Figure 4. The bright, narrow jet (or hot component of the jet) is clearly detected in both the 304~{\AA} (indicated by a blue arrow in Figure 4(c1)) and 131~{\AA} (indicated by a blue arrow in Figure 4(d1)) channels, but is absent in the H$\alpha$ image. This absence suggests that the plasma has been heated to coronal temperatures. The hot component appears intermittently and persists until 02:30:06 UT, as indicated by the blue arrows in Figures 4(d1)-(d4) (see also Animation 2 for additional details). In contrast, the cool component, located to the left of 
the hot jet, is observed as a dark structure in the H$\alpha$ image (indicated by a white arrow in Figure 4(a2)), as well as in the 304~{\AA} and 131~{\AA} channels (indicated by white arrows in Figures 4(c2) and (d2)). 
This indicates that the plasma within the mini-filament was not significantly heated during its eruption. The cool component exhibited adjacent redshifted 
and blueshifted signatures (marked by red and blue arrows in Figures 4(b2) and (b3)), signifying a clockwise rotational motion of the jet. These Doppler shift patterns are highly consistent with the helical signatures reported in high-resolution 
observations of jet activities \citep{2019ApJ...887...56T,2022ApJ...939...25P}. By approximately 02:20:22 UT, the jet reached a maximum projected length of approximately 26 Mm. Thereafter, it gradually descended toward the solar surface, 
as evidenced by redshift signatures in the H$\alpha$ Dopplergrams (indicated by black arrows in Figures 4(b4) and (b5)). The jet completely disappeared after 02:35:09 UT .

As mentioned above, the cool component of the jet was first ejected and later fell back to the solar surface. Both stages were captured by the CHASE spectral observations. Spectra from two specific time points were selected and presented in Figure 5. In the composite H$\alpha$ intensity images, the jet appears as a narrow, dark structure, indicated by white arrows in Figures 5(a) and (d). It is observed as a blue-shifted structure at 02:14:04 UT (indicated by a black arrow in Figure 5(b)) and as a red-shifted structure at 02:21:10 UT (indicated by a black arrow in Figure 5(e)), consistent with the H$\alpha$ Dopplergram results. The average spectra within the red and blue boxes are shown in Figures 5(c) and (f), respectively. The spectrum in Figure 5(c) exhibits a complete blue shift of approximately 7 km s$^{-1}$, whereas the spectrum in Figure 5(f) displays a complete red shift of about 11.4 km s$^{-1}$.

To investigate the kinematics of the jet, we constructed time-distance diagrams using SDO/AIA images at wavelengths of 304~{\AA}, 171~{\AA}, 193~{\AA}, 211~{\AA}, 335~{\AA}, and 131~{\AA} along the slice “A-B" in Figure 1(b), as shown in Figure 6. In these diagrams, the jet appears as a dark feature, indicating the presence of relatively cool plasma. Its characteristic mountain-shaped trajectory clearly reflects both its ascending and descending phases. To quantify the jet’s kinematic properties, we fitted the jet front with a quadratic function s(t)=s$_{0}$+v$_{0}$t+$\frac{1}{2}$at$^2$, where s(t) is the displacement, v$_{0}$ the initial velocity, and a the acceleration. The instantaneous velocity is then given by v(t)=v$_{0}$+at. The average ascending (v$_{up-ave}$) and descending (v$_{down-ave}$) velocities of the jet were calculated by averaging v(t) over their respective temporal intervals. The derived accelerations range from 104.0 to 150.0 m s$^{-2}$ across different wavelengths, exhibiting little variation and remaining well below the solar free-fall value of 274 m s${^{-2}}$. This suggests that the jet dynamics are moderated by additional forces beyond gravity, such as magnetic tension. The average ascending velocities lie between 24.4 to 38.9 km s$^{-1}$, while the average descending velocities range from 29.3 to 38.5 km s$^{-1}$. Overall, the ascending and descending velocities are comparable, indicating that the jet undergoes a relatively symmetric rise and fall. When the line-of-sight velocity is taken into account, the more realistic upward velocity ranges from 25.4 to 39.5 km s$^{-1}$ at 02:14:04 UT, while the corresponding downward velocity ranges from 31.4 to 40.2 km s$^{-1}$ at 02:21:10 UT.

\subsection{Evolution of the Photosphere at the Jet Base}
 
Figures 7(a)-(h) illustrate the temporal evolution of the photospheric magnetic field at the base of the jet. White and black patches represent positive and negative magnetic polarities, respectively. At approximately 00:12:37 UT, an emerging bipole surfaced at the edge of the sunspot (indicated by yellow arrows, see Animation 3). The positive and negative polarities of this bipole are tracked by blue and green arrows in Figures 7(b)–(h) and 7(b)–(d), respectively. These magnetic features exhibit the typical characteristics of Type I MMFs. Over time, both polarities expanded in area and separated. The positive patch drifted northwest, while the negative one migrated southeast until it eventually coalesced with the sunspot. Subsequently, magnetic cancellation took place when the emerging positive polarity collided with a pre-existing negative polarity (marked by the red arrows in Figures 7(a)-(h)). As this process progressed, both the emerging positive and the pre-existing negative polarities gradually diminished in size and intensity. To quantify the motion of the emerging positive polarity, a time-distance diagram was constructed using a sequence of magnetograms along the “C-D" slice in Figure 7(b). The two red dashed lines indicate the start and end times of the jet. The emerging positive polarity exhibited persistent motion toward the negative polarity region, with an average velocity of 333.9$\pm$12.0 m s$^{-1}$. Figure 7(f) presents the temporal evolution of both positive and negative magnetic fluxes within the blue rectangular region outlined in Figure 7(a). The black dashed lines indicate the duration of the jet. Prior to 01:30 UT, the positive flux associated with the emerging positive polarity increased, consistent with the emergence phase. Following 01:30 UT and throughout the jet activity, the flux steadily decreased, signifying ongoing magnetic cancellation. The average magnetic flux-loss rate was estimated to be 3.9$\times$10$^{17}$ Mx hr$^{-1}$. Assuming balanced cancellation between opposite polarities, the corresponding total cancellation rate is estimated to be approximately 7.8$\times$10$^{17}$ Mx hr$^{-1}$. This value is more than an order of magnitude lower than the typical flux-cancellation rates reported for active region jets associated with mini-filament eruptions (typically $\sim$10$^{19}$ Mx hr$^{-1}$) \citep{2017ApJ...844...28S,2025ApJ...994..164P}. Nevertheless, the observed magnetic flux cancellation may have served as the trigger for the jet eruption.

\section{Conclusions and Discussions}

In this study, we present observational evidence for a jet triggered by magnetic reconnection between super-penumbral fibrils and a mini-filament. The event was initiated by the outward migration of an emerging positive polarity associated with a Type I MMF, which acted as a driver to progressively stretch the super-penumbral fibrils. When these fibrils interacted with the magnetic field of the mini-filament, magnetic reconnection was triggered at their interface, as indicated by intense localized brightenings and a peak DEM temperature of approximately 2.5 MK. The reconnection produced a narrow hot jet and a broader cool component, with the cool plasma exhibiting a clockwise rotational motion. CHASE H$\alpha$ spectroscopic observations reveal the jet dynamics, showing blueshifts of $\sim$7 km s$^{-1}$ during the ascending phase and redshifts of $\sim$11.4 km s$^{-1}$ during the descending phase. The coexistence of blueshifted and redshifted components across the jet body provides kinematic evidence for the transfer of magnetic twist from the mini-filament system to the jet during magnetic field reconfiguration and footpoint displacement. Compared with typical active-region jets, this event exhibits lower propagation velocities and lower footpoint temperatures, consistent with a relatively low magnetic flux cancellation rate and indicating a comparatively weak energy release. Magnetic flux cancellation between the emerging positive polarity and the pre-existing negative polarity was observed and is considered the primary driver of the jet eruption. These observations further emphasize the important role of fine-scale magnetic structures, such as MMFs, in triggering magnetic reconnection and driving solar jet activity.

Figure 8 presents a simplified schematic illustrating the jet-generation process. The large gray-filled circles represent the sunspot magnetic field. The small black-filled circles and adjacent white-filled circles denote the emerging bipole of 
the Type I MMF, while the isolated small black-filled circles indicate the pre-existing negative polarity. The “+” and “–” symbols mark the positive and negative magnetic polarities, respectively. Dark purple lines depict the super-penumbral fibrils, 
and orange lines represent the magnetic field of the mini-filament, with one footpoint embedded in the pre-existing negative polarity. The light purple lines show the reconnected magnetic field lines along which the jet propagates, while green lines correspond to the newly formed small reconnected loops. As shown in Figure 8(a), prior to the jet, the two ends of the super-penumbral fibrils were connected to the emerging bipole. The emerging positive polarity moved northwest, while its negative counterpart migrated southeast until it eventually merged into the sunspot. The motion of the emerging positive polarity is indicated by the black arrow. As this positive polarity moved outward, the super-penumbral fibrils became increasingly stretched. When the super-penumbral fibrils approached the magnetic field of the mini-filament, magnetic reconnection took place, as illustrated in Figure 8(b). This reconnection produced a twisted jet ejected along the newly formed reconnected field lines, with one end rooted in the emerging negative polarity within the sunspot (see Figure 8(c)). Simultaneously, a series of small reconnected loops formed between the emerging positive and pre-existing negative polarities, signifying a change in magnetic topology. Consequently, one footpoint of the mini-filament shifted toward the emerging negative polarity. These small reconnected loops eventually submerged below the solar surface, driven by magnetic flux cancellation. Simultaneously, the magnetic twist stored in the mini-filament was transferred to the jet, contributing to the observed rotational motion of the erupting plasma.

A mini-filament was involved in this event, and the overall eruption dynamics are broadly consistent with the classical breakout-jet framework \citep{2015Natur.523..437S}. Our observations suggest that persistent magnetic flux cancellation contributed to the formation and destabilization of the mini-filament, although the complete formation process was not fully captured due to limited observational coverage. The continued flux cancellation near one footpoint of the mini-filament likely promoted the gradual 
buildup of a sheared core field, eventually driving the magnetic system toward an unstable state and resulting in the mini-filament eruption. In the standard breakout model, a sheared magnetic core field hosting the mini-filament is embedded 
beneath a closed overlying magnetic field envelope. As the mini-filament rises, it interacts with the restraining overlying field and drives external (breakout) reconnection at an upper magnetic null point or quasi-separatrix layer, 
thereby weakening the magnetic confinement and facilitating the outward eruption. Meanwhile, internal reconnection beneath the erupting core produces heated plasma and compact closed loops.

In a statistical analysis of 20 jets originating from the peripheries of active regions, \citet{2016A&A...589A..79M} reported propagation velocities in the range of 87-532 km s$^{-1}$, with peak footpoint temperatures reaching Log T = 6.5 (3.0 MK). In contrast, the jet studied here exhibits significantly lower velocities and footpoint temperatures than those reported by \citet{2016A&A...589A..79M}. Given that both studies focus on sunspot peripheries, this discrepancy suggests that the source location is not the primary factor. Instead, it points toward differences in local magnetic topology and reconnection dynamics. The reduced kinematic and thermal properties imply a weaker energy release, likely due to less efficient reconnection or a limited reservoir of free magnetic energy. In addition, the efficiency of energy conversion may be influenced by the flux cancellation rate. For example, \citet{2017ApJ...844...28S} and \citet{2025ApJ...994..164P} reported flux cancellation rates of $\sim$10$^{19}$ Mx hr$^{-1}$ for active-region jets associated with mini-filaments, which are significantly higher than that inferred in our case ($\sim$10$^{17}$ Mx hr$^{-1}$). Instead, our value is comparable to those reported for coronal hole jets \citep{2018ApJ...853..189P}. This difference further supports the interpretation that our event represents a comparatively low-energy jet.

Rotational motion is a ubiquitous feature of solar jets. In a statistical study by \citet{2015ApJ...806...11M}, large coronal jets in coronal holes were found to exhibit an average twist of about 1.5 turns. This magnetic twist is generally transferred from closed magnetic fields to the ambient open field through interchange reconnection, and it manifests spectroscopically as co-spatial blueshifted and redshifted signatures across the jet body. \citet{2015ApJ...801...83C} reported that recurrent coronal jets display helical motions of the same sense in raster Doppler maps, with velocities reaching $\pm$100 km s$^{-1}$. Similarly, \citet{2022ApJ...939...25P} identified opposite Doppler shifts along the two edges of a jet spire using Mg~${\sc II}$ spectra. At smaller scales, jet-like events have also been found to exhibit simultaneous blueshifted and redshifted signatures during their eruption \citep{2018ApJ...869..147T}. In this study, similar Doppler signatures, namely, the coexistence of blueshifted and redshifted components, are detected in the H$\alpha$ Dopplergrams, providing clear evidence of rotational motion in the observed jet. Similar to the cases reported by the cases of \citet{2015ApJ...806...11M}, \citet{2015ApJ...801...83C}, and \citet{2022ApJ...939...25P}, we suggest that the observed twist originates from the eruption of a mini-filament structure.

To clarify the role of the photospheric driver, we compare our results with the 2.5-dimensional MHD simulations of \citet{2013ApJ...777...16Y}, which model chromospheric anemone jets associated with MMFs. This simulation provides a similar triggering mechanism to that of our event, demonstrating that the outward migration of an MMF in the sunspot moat region can lead to magnetic reconnection. However, the interpretation of the rotational signatures differs between the simulation and our observations. \citet{2013ApJ...777...16Y} attributed the rotational signatures to the propagation of shear Alfvén waves, whereas our multi-wavelength analysis suggests that the clockwise rotation observed in this event is more likely associated with the transfer of magnetic twist during reconnection. In this scenario, magnetic helicity stored in the mini-filament system is transported into the jet spire through reconnection-driven restructuring of the magnetic connectivity. By combining these perspectives, our results highlight the important role of MMFs in transporting magnetic energy and helicity from small-scale photospheric structures into the solar atmosphere.

Magnetic flux emergence and cancellation are both considered important processes in the formation and triggering of coronal jets, although their relative contributions remain under debate. In this event, persistent magnetic flux cancellation near one footpoint of the mini-filament was observed before and during the eruption. This cancellation likely contributed to the buildup of magnetic stress and the destabilization of the magnetic structure supporting the mini-filament, suggesting that magnetic flux cancellation played a key role in triggering the jet. Although magnetic flux emergence was also observed during the evolution of the event, it appears to have mainly modified the local magnetic configuration rather than directly triggering the eruption. This scenario differs from jets driven primarily by magnetic flux emergence \citep[e.g.,][]{1995Natur.375...42Y,2005ApJ...635.1299A,2008ApJ...673L.211M}, in which newly emerging magnetic fields reconnect directly with pre-existing ambient fields to produce jet-like outflows. Instead, our event is more consistent with the mini-filament eruption scenario, where flux convergence and cancellation progressively destabilize the magnetic structure supporting the mini-filament, eventually leading to its eruption and the subsequent formation of a coronal jet \citep[e.g.,][]{2016ApJ...832L...7P,2018ApJ...853..189P,2024ApJ...964....7Y,2024MNRAS.528.1094Y}.

\acknowledgments

The authors thank the referee for her/his constructive suggestions and comments, which help to improve this manuscript. We would like to express our sincere gratitude to Prof. Hui Tian for his valuable discussions. 
We are also grateful to the science teams of NVST and SDO for providing the observational data. This study also makes use of data from the CHASE mission, supported by the China National Space Administration. 
This work is supported by the Strategic Priority Research Program of the Chinese Academy of Sciences (Grant No. XDB0560000), the National Key R$\&$D Program of China (No.2024YFA1612001), 
National Natural Science of China (12325303, 12433010, 12273108, 12273060, 12473059, and 12303059), Yunnan Science Foundation of China (202401AT070071 and 202501AT070025), Yunnan Fundamental Research 
Projects (grant NO. 202601AS070076), Youth Innovation Promotion Association, CAS (Nos 2023063), Yunnan Provincial Department of Education Science Research Fund Project (No. 2025J0945),  
and Chuxiong Normal University Doctoral Research Initiation Fund Project (No. BSQD2420).

\appendix

Three supplemental animations corresponding to Figures 2, 4, and 7 are provided. These animations show, respectively, the magnetic reconnection between the super-penumbral fibrils and nearby cool loops, the temporal evolution of the jet across multiple wavelengths, and the evolution of the magnetic field at the jet base.

\clearpage

\begin{figure}
\centering
\includegraphics[scale=1.0]{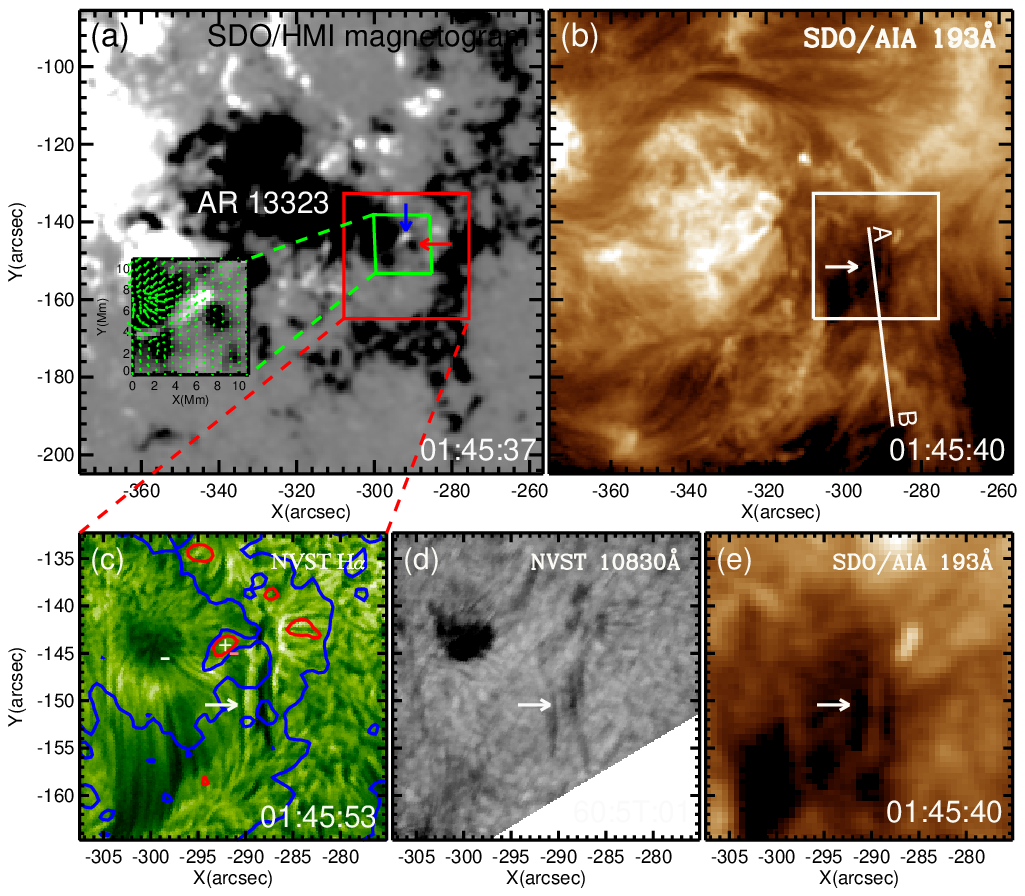}
\caption{Appearance of the active region producing the jet. Panels (a) and (b) show the HMI line-of-sight magnetogram and the SDO/AIA 193~{\AA} image, respectively, indicating the location of the jet. 
Panels (c)-(e) present the magnetic configuration prior to the jet, as observed in NVST H$\alpha$ line-center, NVST He I 10830~{\AA}, and SDO/AIA 193~{\AA} images. The zoomed-in image in panel (a) shows 
the SDO/HMI vertical magnetogram, with the horizontal magnetic field represented by green arrows. Its field of view (FOV) is outlined by the green square. In panel (c), the H$\alpha$ image is 
overlaid with SDO/HMI line-of-sight magnetic field contours, where red and blue contours indicate positive and negative magnetic polarities, respectively, with a contour level of $\pm$50 G. 
The red and white squares in panels (c) and (e) indicate the source region of the jet and the FOV of panels (c)-(d). The white arrows in panels (c)-(e) highlight the mini-filament. 
The ``+'' and ``-'' in panel (c) indicate the positive and the negative magnetic polarities, respectively. \label{fig:fig1}}
\end{figure}

\clearpage

\begin{figure}
\centering
\includegraphics[scale=1.0]{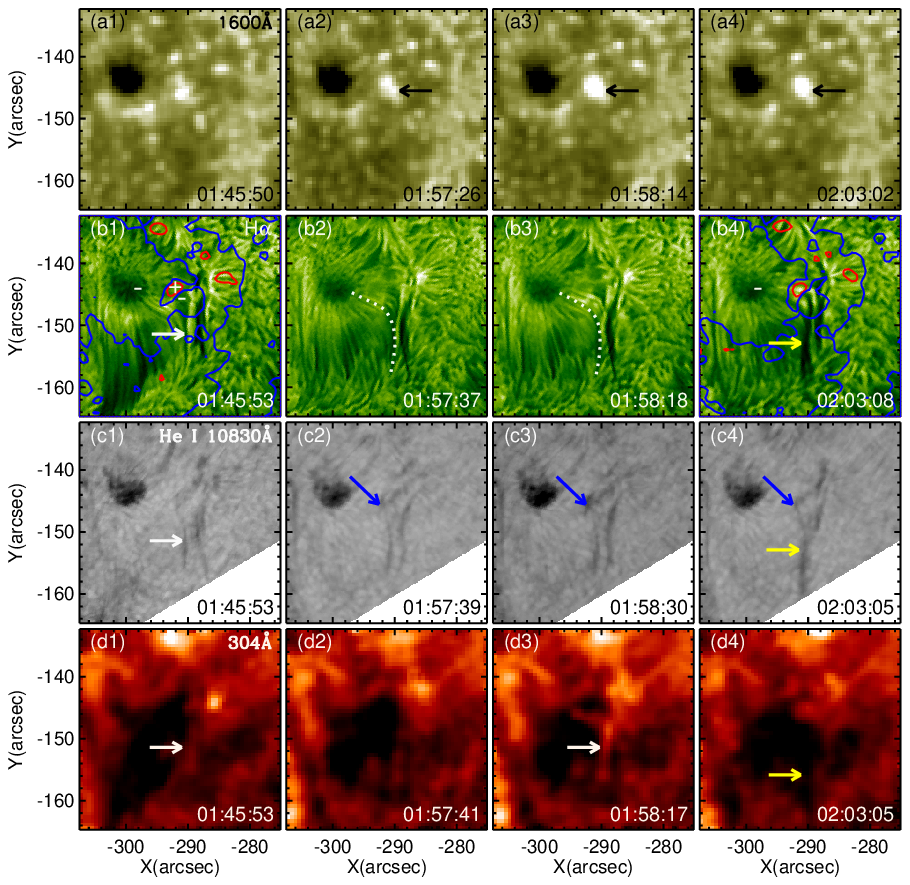}
\caption{Magnetic reconnection process between the super-penumbral fibrils and the mini-filament. Panels (a1) -- (a4), (b1) -- (b4), (c1) -- (c4), and (d1) -- (d4) show this process in SDO/AIA 1600~{\AA}, 
NVST H$\alpha$ line-center, NVST He {\sc i} 10830~{\AA}, and SDO/AIA 304~{\AA} images, respectively. Arrows in panels (a2) -- (a4) and (c2) -- (c3) mark the brightenings at the jet base. The ``+'' and ``-'' in 
panel (c) indicate the positive and the negative magnetic polarities, respectively. In panels (b1) and (b4), the H$\alpha$ images are overlaid with SDO/HMI line-of-sight magnetic field contours, where red and blue 
denote positive and negative polarities, respectively, at a contour level of $\pm$50 G. The “+” and “–” symbols indicate the positive and the negative magnetic polarities, respectively. An online animation of 
SDO/AIA 1600~{\AA}, NVST H$\alpha$ line-center, NVST He I 10830~{\AA} and SDO/AIA 304~{\AA} wavelengths is available. The $\sim$1 s animation covers from $\sim$01:45 to $\sim$02:01 UT.  \label{fig:fig2}}
\end{figure}

\clearpage

\begin{figure}
\centering
\includegraphics[scale=1.0]{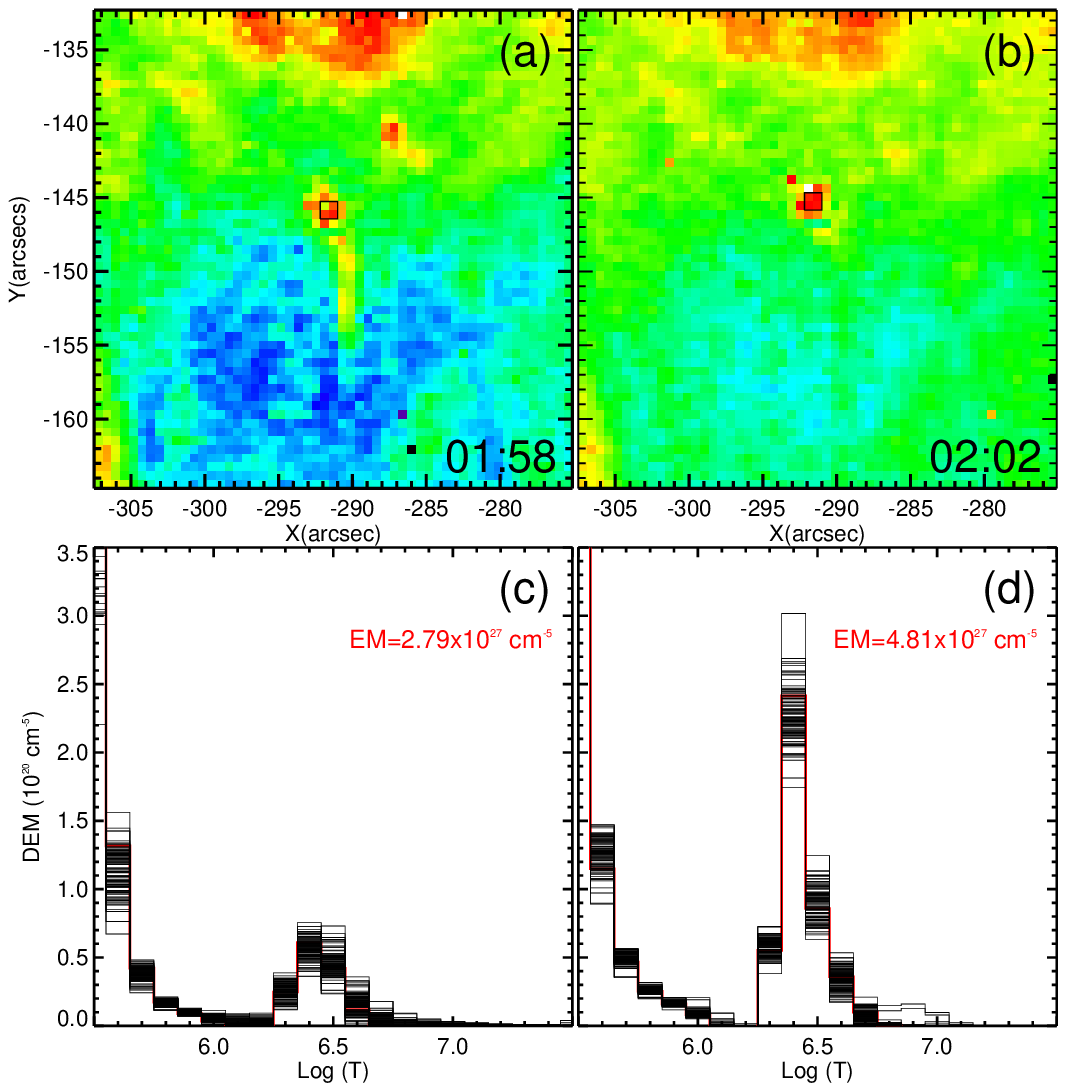}
\caption{Emission measure and DEM diagnostics of jet-associated brightenings. Panels (a) and (b) show the brightenings produced by ongoing magnetic reconnection in the constructed emission measure 
images at 01:58 UT and 02:02 UT, respectively. Panels (c) and (d) present the pre-event-subtracted (01:56 UT) average DEM distributions of the brightenings outlined by the black boxes in panels (a) and (b). The black curves 
represent 100 Monte Carlo simulations, while the red curve denotes the best-fit DEM profile. \label{fig:fig3}}
\end{figure}

\clearpage

\begin{figure}
\centering
\includegraphics[scale=1.0]{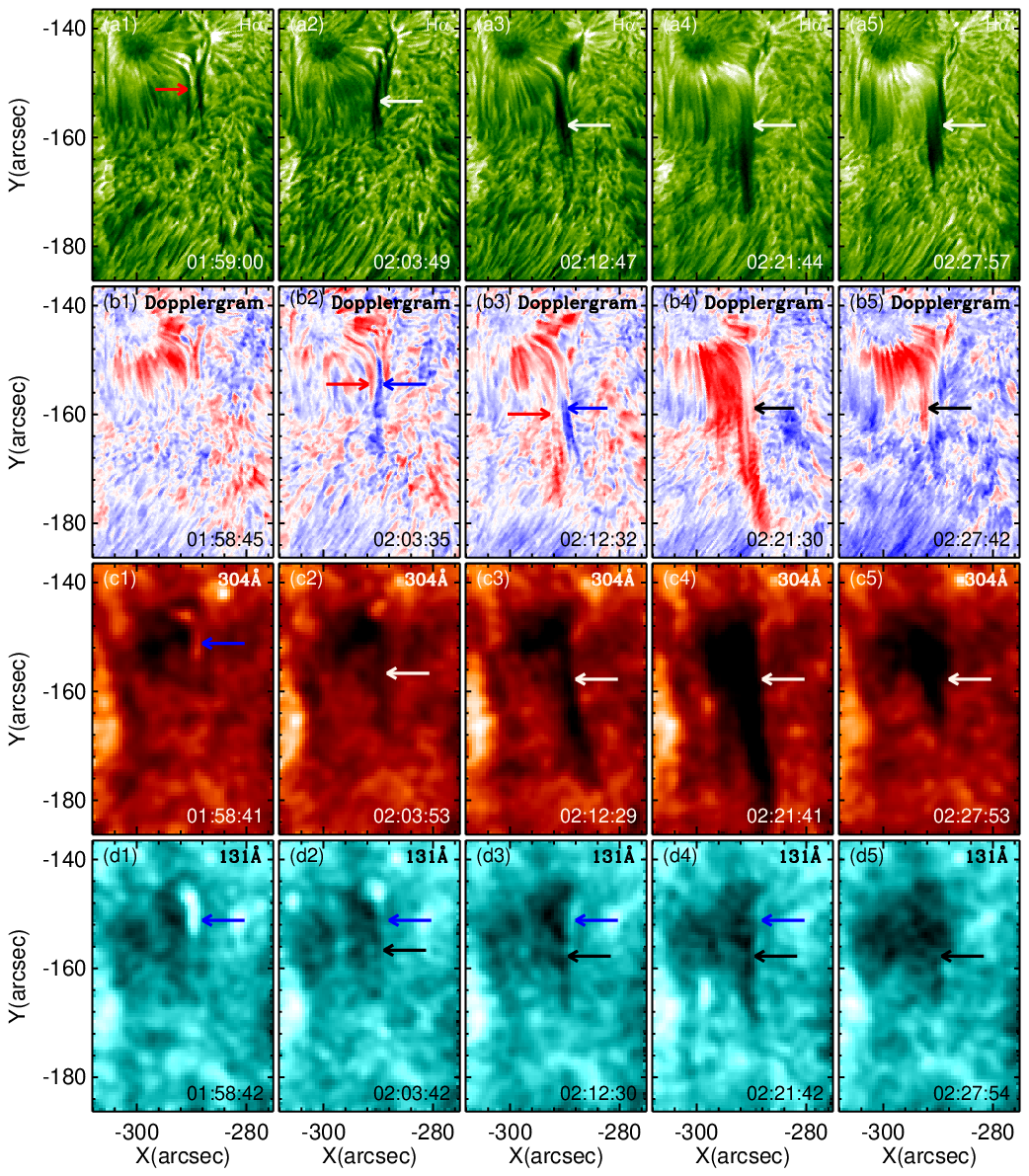}
\caption{The temporal evolution of the jet across multiple wavelengths. Panels (a1)-(a5) show the NVST H$\alpha$ line-center images, where the white arrows mark the primary jet eruption. Panels (b1)-(b5) present the NVST H$\alpha$ Dopplergrams, with red and blue arrows in panels (b2)-(b3) highlighting the adjacent red-shift and blue-shift signatures, while black arrows in panels (b4)-(b5) indicate the subsequent plasma fallback. Panels (c1)-(c5) and (d1)-(d5) display the SDO/AIA 304~{\AA} and 131~{\AA} images, respectively, showing the multi-thermal nature of the jet spire. Blue arrows in (c1) and (d1)-(d4) denote the hot component, whereas white and black arrows highlight the associated cool component. An online animation of NVST H$\alpha$ line-center images, NVST H$\alpha$ Dopplergrams, SDO/AIA 304~{\AA} and 131~{\AA} images is available. The $\sim$4 s animation covers from $\sim$01:45 to $\sim$02:50 UT. \label{fig:fig4}}
\end{figure}

\clearpage

\begin{figure}
\centering
\includegraphics[scale=1.0]{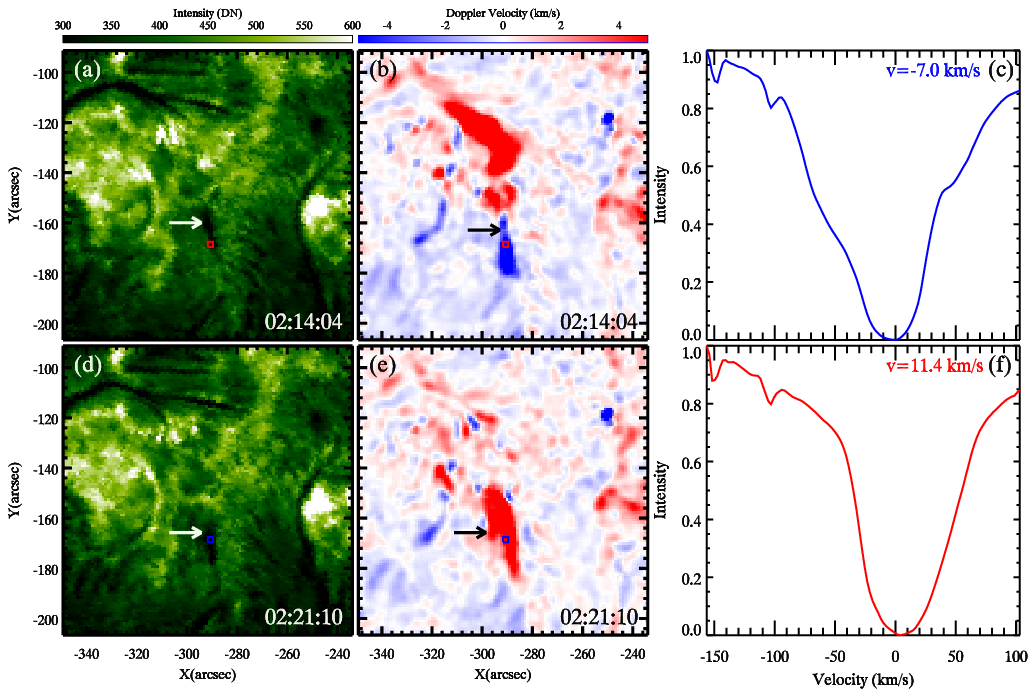}
\caption{CHASE/HIS H$\alpha$ spectral observations. Panels (a) and (b) show the jet in a composite H$\alpha$ line-center intensity image and the corresponding Dopplergram, respectively, derived from scanning spectral data acquired around 02:14 UT. Panel (c) displays the H$\alpha$ line profile extracted from the region outlined by the red boxes in panels (a) and (b). Panels (d) and (e) present the jet in a similar composite H$\alpha$ line-center image and Dopplergram, based on data taken around 02:21 UT. Panel (f) shows the H$\alpha$ line profile from the region marked by the blue boxes in panels (d) and (e). \label{fig:fig5}}
\end{figure}

\clearpage

\begin{figure}
\centering
\includegraphics[scale=1.0]{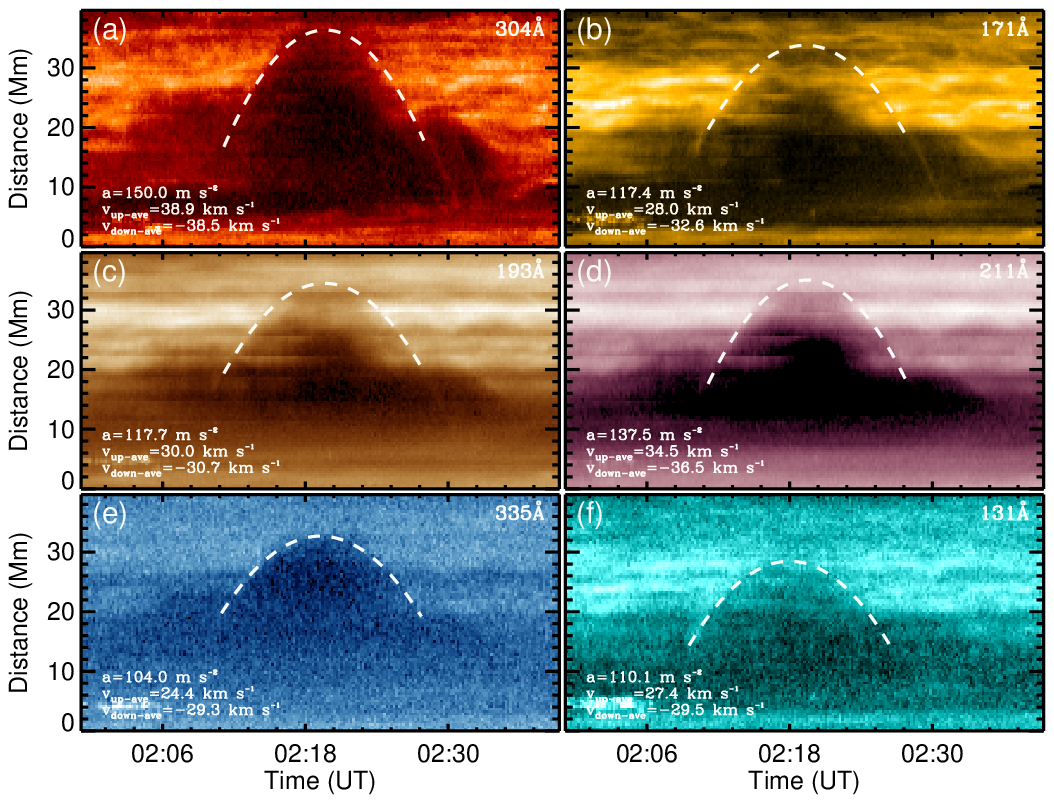}
\caption{Jet kinematics parameter derived from time–distance analysis. Panels (a)–(f) show time–distance diagrams along the line ``A'' -- ``B''  (as indicated in Figure 1(b)) at 304~{\AA}, 171~{\AA}, 193~{\AA}, 211~{\AA}, 335~{\AA}, and 131~{\AA}, respectively. The dashed curve denotes the quadratic fitting result of the jet front. The corresponding acceleration (a), the average upward velocity (v$_{up-ave}$), and the average downward velocity (v$_{down-ave}$) of the jet
are provided in each panel.  \label{fig:fig6}}
\end{figure}

\clearpage

\begin{figure}
\centering
\includegraphics[scale=1.0]{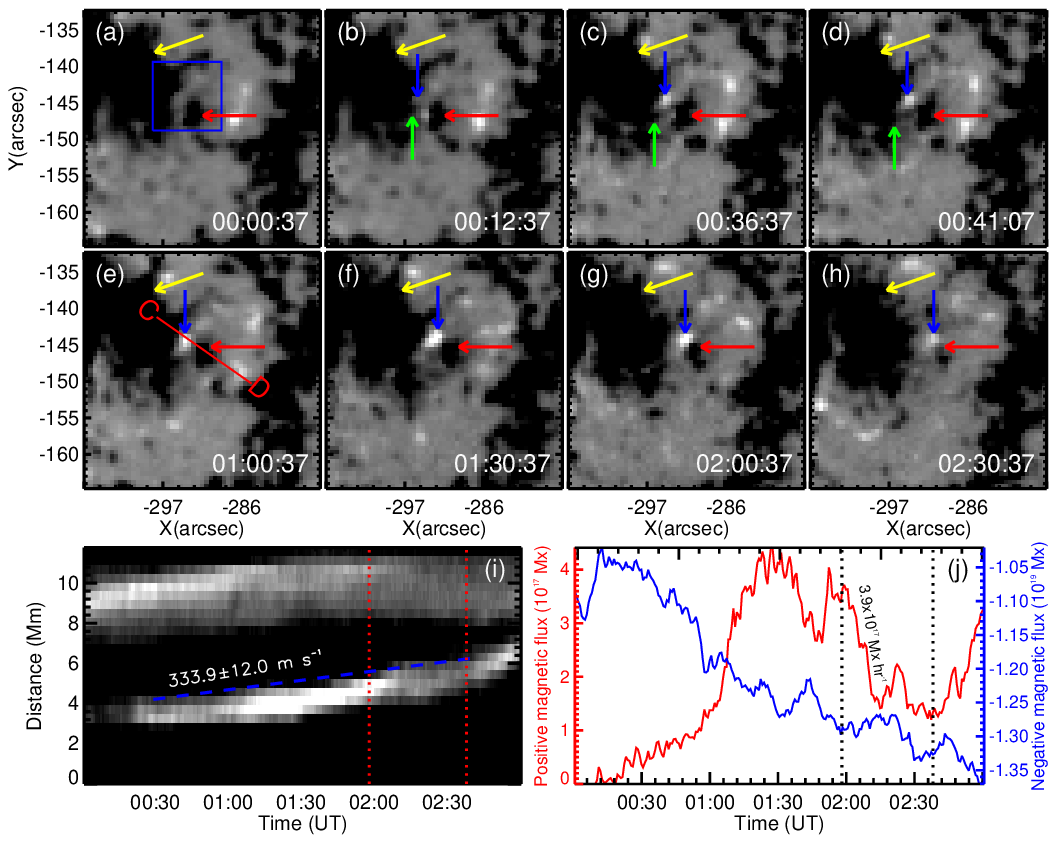}
\caption{Evolution of the magnetic field at the jet base and the corresponding magnetic flux changes. Panels (a)–(h) show a sequence of SDO/HMI line-of-sight magnetograms of the jet base. Panel (i) shows the time–distance diagram along the line ``C'' -- ``D'' indicated in panel (e). Panel (j) presents the variations of positive and negative magnetic fluxes within the blue boxed region outlined in panel (a). The blue and green arrows mark the emerging magnetic bipole of an MMF, while the red arrows denote the pre-existing negative-polarity region. The two red dashed lines mark the start and end times of the jet, respectively, while the two black dashed lines mark the same time interval. An online animation of SDO/HMI line-of-sight magnetograms is available. 
The $\sim$10 s animation covers from $\sim$00:00 to $\sim$02:59 UT. \label{fig:fig7}}
\end{figure}

\clearpage

\begin{figure}
\centering
\includegraphics[scale=0.35]{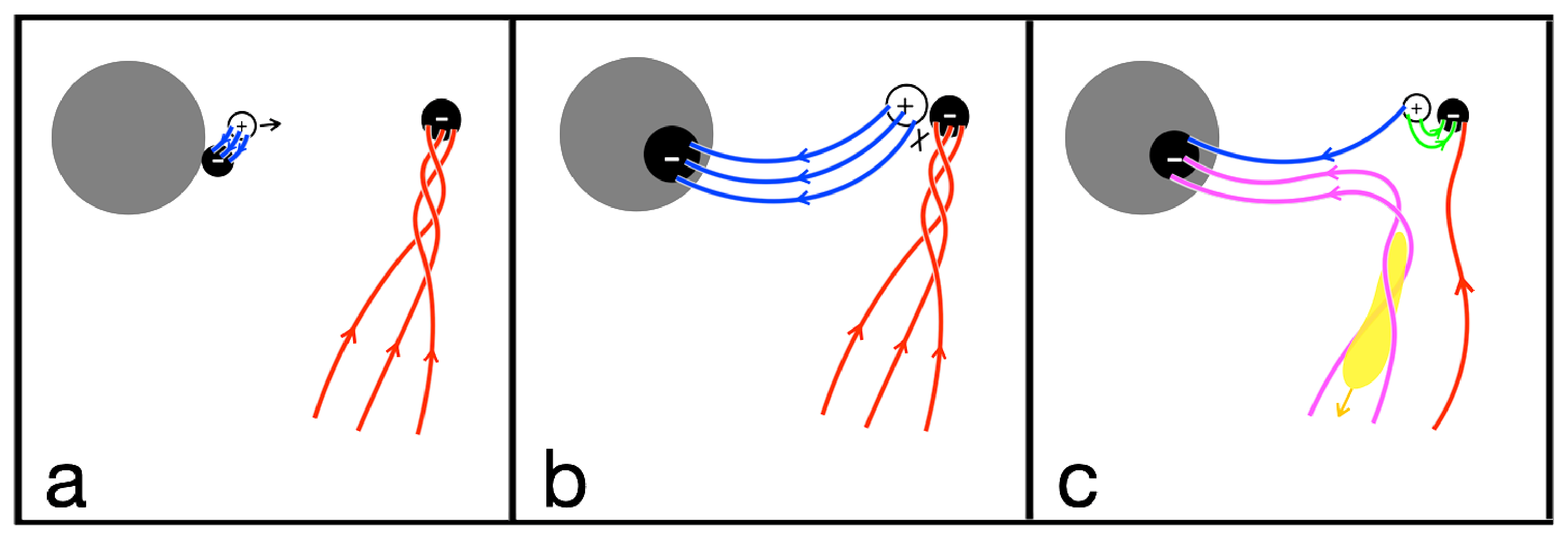}
\caption{Cartoon illustrating the generation process of the jet. (a) Magnetic field topology prior to the jet. The large gray-filled circle denotes the sunspot. The ``+'' and ``--'' symbols denote the positive and the negative magnetic polarities, respectively. The negative and the positive polarities nearby the large circle together form the emerging magnetic bipole of an MMF. The negative polarity outside the large circle represents the pre-existing negative magnetic field. 
The dark purple lines show the super-penumbral fibrils, and the orange lines show the mini-filament. (b) Magnetic reconnection between the super-penumbral fibrils and the mini-filament. The black pentagram marks the reconnection site. 
(c) Magnetic field topology after the jet. The light purple and green lines indicate the newly formed magnetic loops produced by magnetic reconnection. The yellow patch denotes the jet material driven by the reconnection. \label{fig:fig8}}
\end{figure}

\clearpage


\begin{thebibliography}{}

\bibitem[Archontis et al.(2005)]{2005ApJ...635.1299A} Archontis, V., Moreno-Insertis, F., Galsgaard, K., et al.\ 2005, \apj, 635, 2, 1299. doi:10.1086/497533
\bibitem[Brooks et al.(2007)]{2007ApJ...656.1197B} Brooks, D.~H., Kurokawa, H., \& Berger, T.~E.\ 2007, \apj, 656, 2, 1197. doi:10.1086/510144
\bibitem[Cai et al.(2022)]{2022RAA....22f5010C} Cai, Y.-F., Yang, X., Xiang, Y.-Y., et al.\ 2022, Research in Astronomy and Astrophysics, 22, 6, 065010. doi:10.1088/1674-4527/ac69b9
\bibitem[Canfield et al.(1996)]{1996ApJ...464.1016C} Canfield, R.~C., Reardon, K.~P., Leka, K.~D., et al.\ 1996, \apj, 464, 1016. doi:10.1086/177389
\bibitem[Chae et al.(1999)]{1999ApJ...513L..75C} Chae, J., Qiu, J., Wang, H., et al.\ 1999, \apjl, 513, 1, L75. doi:10.1086/311910
\bibitem[Chen et al.(2020)]{2020ApJ...902....8C} Chen, H., Hong, J., Yang, B., et al.\ 2020, \apj, 902, 1, 8. doi:10.3847/1538-4357/abb1c1
\bibitem[Chen et al.(2022)]{2022FrASS...8..238C} Chen, J., Erd{\'e}lyi, R., Liu, J., et al.\ 2022, Frontiers in Astronomy and Space Sciences, 8, 238. doi:10.3389/fspas.2021.786856
\bibitem[Chen et al.(2015)]{2015ApJ...815...71C} Chen, J., Su, J., Yin, Z., et al.\ 2015, \apj, 815, 1, 71. doi:10.1088/0004-637X/815/1/71
\bibitem[Cheng et al.(2012)]{2012ApJ...761...62C} Cheng, X., Zhang, J., Saar, S.~H., et al.\ 2012, \apj, 761, 1, 62. doi:10.1088/0004-637X/761/1/62
\bibitem[Cheung et al.(2015)]{2015ApJ...801...83C} Cheung, M.~C.~M., De Pontieu, B., Tarbell, T.~D., et al.\ 2015, \apj, 801, 2, 83. doi:10.1088/0004-637X/801/2/83
\bibitem[Chitta et al.(2023)]{2023Sci...381..867C} Chitta, L.~P., Zhukov, A.~N., Berghmans, D., et al.\ 2023, Science, 381, 6660, 867. doi:10.1126/science.ade5801
\bibitem[Duan et al.(2024)]{2024ApJ...962L..38D} Duan, Y., Tian, H., Chen, H., et al.\ 2024, \apjl, 962, 2, L38. doi:10.3847/2041-8213/ad24f3
\bibitem[Guo et al.(2013)]{2013A&A...555A..19G} Guo, Y., D{\'e}moulin, P., Schmieder, B., et al.\ 2013, \aap, 555, A19. doi:10.1051/0004-6361/201321229
\bibitem[Hinode Review Team et al.(2019)]{2019PASJ...71R...1H} Hinode Review Team, Al-Janabi, K., Antolin, P., et al.\ 2019, \pasj, 71, 5, R1. doi:10.1093/pasj/psz084
\bibitem[Hong et al.(2017)]{2017ApJ...835...35H} Hong, J., Jiang, Y., Yang, J., et al.\ 2017, \apj, 835, 1, 35. doi:10.3847/1538-4357/835/1/35
\bibitem[Hou et al.(2020)]{2020A&A...642A..44H} Hou, Y.~J., Li, T., Zhong, S.~H., et al.\ 2020, \aap, 642, A44. doi:10.1051/0004-6361/202038668
\bibitem[Innes et al.(2016)]{2016AN....337.1024I} Innes, D.~E., Bu{\v{c}}{\'\i}k, R., Guo, L.-J., et al.\ 2016, Astronomische Nachrichten, 337, 10, 1024. doi:10.1002/asna.201612428
\bibitem[Joshi et al.(2017)]{2017Ap&SS.362...10J} Joshi, N.~C., Chandra, R., Guo, Y., et al.\ 2017, \apss, 362, 1, 10. doi:10.1007/s10509-016-2983-x
\bibitem[Jiang et al.(2007)]{2007A&A...469..331J} Jiang, Y.~C., Chen, H.~D., Li, K.~J., et al.\ 2007, \aap, 469, 1, 331. doi:10.1051/0004-6361:20053954
\bibitem[Jing et al.(2019)]{2019ApJ...880..143J} Jing, J., Li, Q., Liu, C., et al.\ 2019, \apj, 880, 2, 143. doi:10.3847/1538-4357/ab2b44
\bibitem[Lee et al.(2015)]{2015ApJ...798L..10L} Lee, E.~J., Archontis, V., \& Hood, A.~W.\ 2015, \apjl, 798, 1, L10. doi:10.1088/2041-8205/798/1/L10
\bibitem[Lemen et al.(2012)]{2012SoPh..275...17L} Lemen, J.~R., Title, A.~M., Akin, D.~J., et al.\ 2012, \solphys, 275, 1-2, 17. doi:10.1007/s11207-011-9776-8
\bibitem[Li et al.(2022)]{2022SCPMA..6589602L} Li, C., Fang, C., Li, Z., et al.\ 2022, Science China Physics, Mechanics, and Astronomy, 65, 8, 289602. doi:10.1007/s11433-022-1893-3
\bibitem[Li et al.(2019)]{2019ApJ...876..129L} Li, Q., Deng, N., Jing, J., et al.\ 2019, \apj, 876, 2, 129. doi:10.3847/1538-4357/ab18aa
\bibitem[Li et al.(2018)]{2018NatSR...8.8136L} Li, X., Zhang, J., Yang, S., et al.\ 2018, Scientific Reports, 8, 8136. doi:10.1038/s41598-018-26581-4
\bibitem[Li et al.(2023)]{2023ApJ...947L..17L} Li, X., Keppens, R., \& Zhou, Y.\ 2023, \apjl, 947, 1, L17. doi:10.3847/2041-8213/acc9ba
\bibitem[Liu et al.(2011)]{2011ApJ...735L..18L} Liu, C., Deng, N., Liu, R., et al.\ 2011, \apjl, 735, 1, L18. doi:10.1088/2041-8205/735/1/L18
\bibitem[Liu et al.(2022)]{2022RAA....22i5005L} Liu, H., Jin, Z., Xiang, Y., et al.\ 2022, Research in Astronomy and Astrophysics, 22, 9, 095005. doi:10.1088/1674-4527/ac7cba
\bibitem[Liu et al.(2016)]{2016ApJ...833..150L} Liu, J., Wang, Y., Erd{\'e}lyi, R., et al.\ 2016, \apj, 833, 2, 150. doi:10.3847/1538-4357/833/2/150
\bibitem[Liu et al.(2014)]{2014RAA....14..705L} Liu, Z., Xu, J., Gu, B.-Z., et al.\ 2014, Research in Astronomy and Astrophysics, 14, 6, 705-718. doi:10.1088/1674-4527/14/6/009
\bibitem[Loughhead(1968)]{1968SoPh....5..489L} Loughhead, R.~E.\ 1968, \solphys, 5, 4, 489. doi:10.1007/BF00147015
\bibitem[Lu et al.(2019)]{2019ApJ...887..154L} Lu, L., Feng, L., Li, Y., et al.\ 2019, \apj, 887, 2, 154. doi:10.3847/1538-4357/ab530c
\bibitem[Miao et al.(2018)]{2018ApJ...869...39M} Miao, Y., Liu, Y., Li, H.~B., et al.\ 2018, \apj, 869, 1, 39. doi:10.3847/1538-4357/aaeac1
\bibitem[Moreno-Insertis et al.(2008)]{2008ApJ...673L.211M} Moreno-Insertis, F., Galsgaard, K., \& Ugarte-Urra, I.\ 2008, \apjl, 673, 2, L211. doi:10.1086/527560
\bibitem[Moore et al.(2010)]{2010ApJ...720..757M} Moore, R.~L., Cirtain, J.~W., Sterling, A.~C., et al.\ 2010, \apj, 720, 1, 757. doi:10.1088/0004-637X/720/1/757
\bibitem[Moore et al.(2015)]{2015ApJ...806...11M} Moore, R.~L., Sterling, A.~C., \& Falconer, D.~A.\ 2015, \apj, 806, 1, 11. doi:10.1088/0004-637X/806/1/11
\bibitem[Mulay et al.(2016)]{2016A&A...589A..79M} Mulay, S.~M., Tripathi, D., Del Zanna, G., et al.\ 2016, \aap, 589, A79. doi:10.1051/0004-6361/201527473
\bibitem[Pariat et al.(2009)]{2009ApJ...691...61P} Pariat, E., Antiochos, S.~K., \& DeVore, C.~R.\ 2009, \apj, 691, 1, 61. doi:10.1088/0004-637X/691/1/61
\bibitem[Panesar et al.(2016)]{2016ApJ...832L...7P} Panesar, N.~K., Sterling, A.~C., Moore, R.~L., et al.\ 2016, \apjl, 832, 1, L7. doi:10.3847/2041-8205/832/1/L7
\bibitem[Panesar et al.(2018)]{2018ApJ...853..189P} Panesar, N.~K., Sterling, A.~C., \& Moore, R.~L.\ 2018, \apj, 853, 2, 189. doi:10.3847/1538-4357/aaa3e9
\bibitem[Panesar et al.(2022)]{2022ApJ...939...25P} Panesar, N.~K., Tiwari, S.~K., Moore, R.~L., et al.\ 2022, \apj, 939, 1, 25. doi:10.3847/1538-4357/ac8d65
\bibitem[Panesar et al.(2025)]{2025ApJ...994..164P} Panesar, N.~K., Sterling, A.~C., Moore, R.~L., et al.\ 2025, \apj, 994, 2, 164. doi:10.3847/1538-4357/ae0d90
\bibitem[Patsourakos \& Archontis(2025)]{2025A&A...699A..87P} Patsourakos, S. \& Archontis, V.\ 2025, \aap, 699, A87. doi:10.1051/0004-6361/202554580
\bibitem[Pesnell et al.(2012)]{2012SoPh..275....3P} Pesnell, W.~D., Thompson, B.~J., \& Chamberlin, P.~C.\ 2012, \solphys, 275, 1-2, 3. doi:10.1007/s11207-011-9841-3
\bibitem[Pontin et al.(2024)]{2024ApJ...960...51P} Pontin, D.~I., Priest, E.~R., Chitta, L.~P., et al.\ 2024, \apj, 960, 1, 51. doi:10.3847/1538-4357/ad03eb
\bibitem[Qi et al.(2022)]{2022A&A...657A.118Q} Qi, Y., Huang, Z., Xia, L., et al.\ 2022, \aap, 657, A118. doi:10.1051/0004-6361/202141401
\bibitem[Qiu et al.(2022)]{2022SCPMA..6589603Q} Qiu, Y., Rao, S., Li, C., et al.\ 2022, Science China Physics, Mechanics, and Astronomy, 65, 8, 289603. doi:10.1007/s11433-022-1900-5
\bibitem[Rao et al.(2016)]{2016ApJ...833..210R} Rao, C., Zhu, L., Rao, X., et al.\ 2016, \apj, 833, 2, 210. doi:10.3847/1538-4357/833/2/210
\bibitem[Raouafi et al.(2016)]{2016SSRv..201....1R} Raouafi, N.~E., Patsourakos, S., Pariat, E., et al.\ 2016, \ssr, 201, 1-4, 1. doi:10.1007/s11214-016-0260-5
\bibitem[Scherrer et al.(2012)]{2012SoPh..275..207S} Scherrer, P.~H., Schou, J., Bush, R.~I., et al.\ 2012, \solphys, 275, 1-2, 207. doi:10.1007/s11207-011-9834-2
\bibitem[Schmieder et al.(2022)]{2022AdSpR..70.1580S} Schmieder, B., Joshi, R., \& Chandra, R.\ 2022, Advances in Space Research, 70, 6, 1580. doi:10.1016/j.asr.2021.12.013
\bibitem[Shibata et al.(1992)]{1992PASJ...44L.173S} Shibata, K., Ishido, Y., Acton, L.~W., et al.\ 1992, \pasj, 44, L173. 
\bibitem[Shibata et al.(2007)]{2007Sci...318.1591S} Shibata, K., Nakamura, T., Matsumoto, T., et al.\ 2007, Science, 318, 5856, 1591. doi:10.1126/science.1146708
\bibitem[Shine \& Title(2000)]{2000eaa..bookE2038S} Shine, R. \& Title, A.\ 2000, Encyclopedia of Astronomy and Astrophysics, 2038. doi:10.1888/0333750888/2038
\bibitem[Sterling et al.(2015)]{2015Natur.523..437S} Sterling, A.~C., Moore, R.~L., Falconer, D.~A., et al.\ 2015, \nat, 523, 7561, 437. doi:10.1038/nature14556
\bibitem[Sterling et al.(2017)]{2017ApJ...844...28S} Sterling, A.~C., Moore, R.~L., Falconer, D.~A., et al.\ 2017, \apj, 844, 1, 28. doi:10.3847/1538-4357/aa7945
\bibitem[Sterling et al.(2024)]{2024ApJ...960..109S} Sterling, A.~C., Moore, R.~L., \& Panesar, N.~K.\ 2024, \apj, 960, 2, 109. doi:10.3847/1538-4357/acff6b
\bibitem[Solanki(2003)]{2003A&ARv..11..153S} Solanki, S.~K.\ 2003, \aapr, 11, 2-3, 153. doi:10.1007/s00159-003-0018-4
\bibitem[Tian et al.(2014)]{2014Sci...346A.315T} Tian, H., DeLuca, E.~E., Cranmer, S.~R., et al.\ 2014, Science, 346, 6207, 1255711. doi:10.1126/science.1255711
\bibitem[Tian et al.(2018)]{2018ApJ...854...92T} Tian, H., Yurchyshyn, V., Peter, H., et al.\ 2018, \apj, 854, 2, 92. doi:10.3847/1538-4357/aaa89d
\bibitem[Tiwari et al.(2018)]{2018ApJ...869..147T} Tiwari, S.~K., Moore, R.~L., De Pontieu, B., et al.\ 2018, \apj, 869, 2, 147. doi:10.3847/1538-4357/aaf1b8
\bibitem[Tiwari et al.(2019)]{2019ApJ...887...56T} Tiwari, S.~K., Panesar, N.~K., Moore, R.~L., et al.\ 2019, \apj, 887, 1, 56. doi:10.3847/1538-4357/ab54c1
\bibitem[Wang \& Sheeley(2002)]{2002ApJ...575..542W} Wang, Y.-M. \& Sheeley, N.~R.\ 2002, \apj, 575, 1, 542. doi:10.1086/341145
\bibitem[Wiegelmann(2004)]{2004SoPh..219...87W} Wiegelmann, T.\ 2004, \solphys, 219, 87. doi:10.1023/B:SOLA.0000021799.39465.36
\bibitem[Wiegelmann et al.(2006)]{2006SoPh..233..215W} Wiegelmann, T., Inhester, B., \& Sakurai, T.\ 2006, \solphys, 233, 215. doi:10.1007/s11207-006-2092-z
\bibitem[Wyper et al.(2017)]{2017Natur.544..452W} Wyper, P.~F., Antiochos, S.~K., \& DeVore, C.~R.\ 2017, \nat, 544, 7651, 452. doi:10.1038/nature22050
\bibitem[Xiang et al.(2016)]{2016NewA...49....8X} Xiang, Y.-. yuan ., Liu, Z., \& Jin, Z.-. yu .\ 2016, \na, 49, 8. doi:10.1016/j.newast.2016.05.002
\bibitem[Yan et al.(2020)]{2020ScChE..63.1656Y} Yan, X., Liu, Z., Zhang, J., et al.\ 2020, Science in China E: Technological Sciences, 63, 9, 1656. doi:10.1007/s11431-019-1463-6
\bibitem[Yang et al.(2019)]{2019ApJ...887..220Y} Yang, B., Yang, J., Bi, Y., et al.\ 2019, \apj, 887, 2, 220. doi:10.3847/1538-4357/ab557e
\bibitem[Yang et al.(2023)]{2023ApJ...942...86Y} Yang, J., Hong, J., Yang, B., et al.\ 2023, \apj, 942, 2, 86. doi:10.3847/1538-4357/aca66f
\bibitem[Yang et al.(2024a)]{2024ApJ...964....7Y} Yang, J., Chen, H., Hong, J., et al.\ 2024a, \apj, 964, 1, 7. doi:10.3847/1538-4357/ad23e5
\bibitem[Yang et al.(2011)]{2011RAA....11.1229Y} Yang, L.-H., Jiang, Y.-C., Yang, J.-Y., et al.\ 2011, Research in Astronomy and Astrophysics, 11, 10, 1229. doi:10.1088/1674-4527/11/10/010
\bibitem[Yang et al.(2019)]{2019ApJ...887..239Y} Yang, L., Yan, X., Xue, Z., et al.\ 2019, \apj, 887, 2, 239. doi:10.3847/1538-4357/ab55d7
\bibitem[Yang et al.(2023)]{2023ApJ...945...96Y} Yang, L., Yan, X., Xue, Z., et al.\ 2023, \apj, 945, 2, 96. doi:10.3847/1538-4357/acb6f6
\bibitem[Yang et al.(2024b)]{2024MNRAS.528.1094Y} Yang, L., Yan, X., Xue, Z., et al.\ 2024b, \mnras, 528, 1, 1094. doi:10.1093/mnras/stad3876
\bibitem[Yang et al.(2025)]{2025ApJ...987..193Y} Yang, L., Yan, X., Zhang, J., et al.\ 2025, \apj, 987, 2, 193. doi:10.3847/1538-4357/addac1
\bibitem[Yang et al.(2013)]{2013ApJ...777...16Y} Yang, L., He, J., Peter, H., et al.\ 2013, \apj, 777, 1, 16. doi:10.1088/0004-637X/777/1/16
\bibitem[Yuan et al.(2019)]{2019ApJ...884L..51Y} Yuan, D., Shen, Y., Liu, Y., et al.\ 2019, \apjl, 884, 2, L51. doi:10.3847/2041-8213/ab4bcd
\bibitem[Yokoyama \& Shibata(1995)]{1995Natur.375...42Y} Yokoyama, T. \& Shibata, K.\ 1995, \nat, 375, 42. doi:10.1038/375042a0
\bibitem[Zhang et al.(2016)]{2016SPIE.9909E..2CZ} Zhang, L., Kong, L., Bao, H., et al.\ 2016, \procspie, 9909, 99092C. doi:10.1117/12.2231955
\bibitem[Zhang et al.(2023)]{2023SCPMA..6669611Z} Zhang, L., Bao, H., Rao, X., et al.\ 2023, Science China Physics, Mechanics, and Astronomy, 66, 6, 269611. doi:10.1007/s11433-022-2107-4
\bibitem[Zhang \& Ji(2014)]{2014A&A...567A..11Z} Zhang, Q.~M. \& Ji, H.~S.\ 2014, \aap, 567, A11. doi:10.1051/0004-6361/201423698
\bibitem[Zhang \& Ni(2019)]{2019ApJ...870..113Z} Zhang, Q.~M. \& Ni, L.\ 2019, \apj, 870, 2, 113. doi:10.3847/1538-4357/aaf391

\end{thebibliography}
\end{document}